\documentclass[preprint,showpacs,preprintnumbers]{revtex4}
\usepackage{graphicx}
\usepackage{dcolumn}
\usepackage{bm}
\usepackage{subfigure}

\newcommand{\be}{\begin{equation}}
\newcommand{\en}{\end{equation}}
\newcommand{\bea}{\begin{eqnarray}}
\newcommand{\ena}{\end{eqnarray}}
\begin{document}

\title{ Reconstructing braneworld inflation}
\author{Ram\'on Herrera}
\email{ramon.herrera@ucv.cl}
\affiliation{Instituto de F\'{\i}sica, Pontificia Universidad Cat\'{o}lica de Valpara%
\'{\i}so, Casilla 4059, Valpara\'{\i}so, Chile.}

\begin{abstract}

The reconstruction of a  braneworld inflationary universe
 considering the parametrization  (or attractor) of   the scalar spectral index $n_s(N)$
   in terms  of the number of $e$-foldings N is developed. We also study
  the possibility that the reconstruction for the scenario  of braneworld
  inflation,
  can be realized in terms of the tensor to scalar ratio $r(N)$.
    For both reconstruction methodologies,  we consider
    a general formalism in order to obtain the effective potential as a function
     of the cosmological parameters $n_s(N)$ or $r(N)$.
   For both  reconstruction methods, we consider the
   specific examples for large $N$ in the framework of  the slow roll approximation as; 
    the attractor  $n_s-1\propto N^{-1}$ for the scalar spectral index
   and the attractor  $r\propto N^{-2}$ for the tensor to scalar ratio.
   In this context and depending on the attractors used,  we find different  expressions
    for the effective potential
   $V(\phi)$,  as also the constraints on the parameters present in the reconstruction.

\end{abstract}

\pacs{98.80.Cq}
\maketitle



\section{Introduction}

It is well known that during the early universe, the introduction
of the  inflationary stage or inflation, is to date a possible
solution to many long-standing problems of the hot big bang model
(horizon, flatness, monopoles, etc.)\cite{,Staro,guth,infla}.
However, the most significant characteristic  of the inflationary
model is that inflation gives account of  a causal interpretation
of the origin of the observed anisotropy of the cosmic microwave
background radiation (CMB), as also the distribution of a large
scale structure observed today\cite{A1,astro,astro2,Planck2018}.

In order to describe the inflationary epoch for the early universe, different
inflationary
models have been proposed in the framework of General Relativity (GR) as in
modified gravity or an  alternative to Einstein's General Relativity. In this
context,
implications of string/M-theory to
Friedmann-Robertson-Walker (FRW) cosmological models have
 attracted a great deal of attention in the last years and in particular  some
 models with brane-antibrane configurations like some 
time-like branes, together with their  applications  to the
inflationary cosmology\cite{sen1}. In this framework,  the
introduction  of extra dimensions  generates  extra terms in the
Friedmann equation product of the dimensional reduction (embedded) to
four-dimensions \cite{1,3,8} and
 the standard model of particles is confined to
the brane, while  the gravitation propagates into the bulk space-time\cite{3}.
In this respect,
the  inflationary model of a 
Randall-Sundrum (RS) type II scenario has taken great attentiveness
in the last years\cite{RS} and
 this modification to GR for the
cosmological models has been widely studied. In particular the chaotic
model on the brane in the framework of slow roll was analyzed in
Ref.\cite{Maartens:1999hf}. In Ref.\cite{Huey:2001ae} an inverse power law potential was
 studied, where
 a single scalar field can act as an inflaton field and quintessence for an
appropriate value of the brane tension. In the case of a tachyonic potential considering
 the power law inflation in the frame of braneworld cosmology was developed in
Ref.\cite{Sami:2002zy}.
For a
comprehensible  review of brane-cosmology, see e.g.
Refs.\cite{4,5,M1} and recent articles;  see the list in \cite{Kallosh:2018zsi}.

On the other hand, the reconstruction of the background and in
particular the  effective potential associated with a  scalar field in
the context of inflation from observational data such as the
scalar spectrum,  scalar spectral index $n_s$ and the tensor to
scalar ratio $r$,   has  been analyzed by several authors
\cite{Hodges:1990bf,HL1,H2,H3,H4,Chiba:2015zpa,H5}. Originally,
considering   a single scalar as 
 the reconstruction of inflationary potentials from the primordial scalar
 spectrum  was proposed
   in Ref.\cite{Hodges:1990bf}.

An attractive   mechanism  to reconstruct the effective
potential of the scalar field assuming   the slow roll approximation is
through the parametrization in terms of the number of $e$-folds $N$. In this
respect,
by considering   the scalar
spectral index $n_s(N)$ and the tensor to scalar ratio $r(N)$ (commonly called attractors) it is possible to
 reconstruct the background during the inflationary epoch.
From an  observational point of view, the attractors given by $n_s-1\propto N^{-1}$ and $r\propto N^{-2}$, by
considering the number of $e$-foldings  $N\simeq 50-70$ at the
end of the inflationary epoch,  agree  with the Planck  results\cite{Planck2018}.
In particular  and considering the framework of GR the scalar spectral index given
 by $n_s(N)- 1\propto N^{-1}$,
it is possible to build different effective potentials such as; the
T-model \cite{T}, E-model\cite{E}, Staronbisky
$R^2$-model\cite{Staro}, the chaotic model\cite{Linde83} and the
model of Higgs inflation with non minimal
coupling\cite{Higgs,Higgs2}. In the framework of warm inflation unlike cold inflation, it  was necessary to
consider jointly the attractors $n_s(N)$ and $r(N)$, in order to
reconstruct the effective potential and the dissipation coefficient \cite{Herrera:2018cgi}.
Analogously, the reconstruction of an inflationary model in the context of the Galileon model or G-model,
considering as attractors the scalar spectral index
 and the tensor to scalar ratio
as a function of the number of $e$-folding jointly was studied in
Ref.\cite{Herrera:2018kera}.

We also mentioned 
 another way to reconstruct the background and it is
related with
 the slow-roll
parameter $\epsilon$ and its parametrization in terms of $N$ i.e.,
$\epsilon(N)$. In this sense, considering $\epsilon(N)$ is
possible to find the scalar spectral index and the tensor to
scalar ratio for inflationary models in GR, see
\cite{Huang:2007qz,Gao:2017owg}. In particular choosing  different
slow-roll parameters $\epsilon(N)$,  the reconstruction of several
effective potentials associated with a  scalar field and the
observational parameters were studied in   Ref.\cite{M1}. Also, in
Refs.\cite{Roest:2013fha,N1,N2} were found the effective potential
and the consistency relation $r=r(n_s)$, but considering the
 two slow roll parameters $\epsilon(N)$ and $\eta(N)$.

 In the context of modified gravity, in
Ref.\cite{Odintsov:2018zhw} the reconstruction of  the effective
potential and the coupling of the Gauss-Bonnet function
  was obtained during the inflationary epoch by fixing the
tensor-to-scalar ratio and the Hubble parameter as a function of
the $e$-folds,  in the framework of Einstein Gauss-Bonnet gravity.
Also, the reconstruction of an  inflationary stage assuming the slow-roll approximation
 for $F(R)$ gravity considering different
expressions for the tensor to scalar ratio in terms of $N$, was
developed in Ref.\cite{Odintsov:2017fnc}, see also
Ref.\cite{Odintsov:2018ggm} for other modified gravities.

On the other hand, the reconstruction technique from different
 Equation of State (EoS) in
the context of the fluid cosmology during inflation   was studied  in Refs.\cite{fluido,fluidoA,expo}
and for the case of the current universe in \cite{fluido2}. In particular for
the reconstruction of the
inflationary epoch it is possible to assume an ansatz on the effective EoS as
function of the number of $e-$foldings\cite{fluido}. Here, rewriting the scalar spectral index
and the tensor to scalar ratio in terms of the effective EoS, one can find the
attractors $n_s(N)$ and $r(N)$ in the fluid inflation. Subsequently, if
the fluid corresponds to a standard scalar field, one can obtain the
reconstruction of the effective potential under slow roll approximation\cite{fluido}.

It is interesting to mention that from the point of view of the
fluid cosmology, it is possible to consider effects of viscosity
terms dependent on the Hubble rate and its derivatives in the EoS
of the
 dark fluid  and then the equations of motion from this fluid can be visualized as modifications to
 the GR, which is how it happens
in some braneworld  models or fourth order gravity, see
Ref.\cite{Capozziello:2005pa}.

Another methodology that has been
 widely studied in the literature for the
reconstruction of the effective potential and the observables $n_s(N)$ and $r(N)$
in the framework of inflation is to
consider the scale factor as ansatz. In this sense, we mentioned the
inflationary models
such as; power-law\cite{powerl}, intermediate\cite{interm}, logamediate\cite{loga},
exponential \cite{expo,Herrera:2018ker}, among others.  Models of dark energy
and its reconstruction from the scale factor were studied in Refs.\cite{m1,m2}.

The goal of this study is to reconstruct the braneworld inflation,
through
 the parametrization of the scalar spectral index or  the tensor
to scalar ratio, as function of the number of e-foldings. In fact,
 we analyze how the brane model changes  the
reconstruction of the scalar potential, considering as attractors
the spectral index $n_s(N)$ or the tensor to scalar ratio $r(N)$.
 In this respect, we will consider the domination of the
brane effect, in order to obtain analytical solutions in the
reconstruction of the background. We will also formulate a general
formalism to find the effective potential, by assuming the
parametrization $n_s(N)$ or $r(N)$, in the context of the slow
roll approximation. Thus,  choosing a specific attractor for the
observable $n_s$ or the tensor to scalar ratio $r$ in terms of the
number of $e$-folds for large $N$, we will show the possibility of 
reconstructing  the effective potential $V(\phi)$, in the frame of
braneworld inflation.

As an application of
the formulated formalism, we will analyze two different
reconstructions. Following the standard reconstruction of the
background from $n_s(N)$, we shall consider the specific case in which the
scalar spectral index is given by $n_s=1-2/N$. As a second
reconstruction, we shall regard the reconstruction from the point
of view of the tensor to scalar ratio $r(N)$ and as it modifies
the reconstruction of the effective potential. In these
reconstructions, we will derive different constraints on the
parameters present in the models.

The outline of the paper is as follows. The next section presents
a brief review of the background and  the
cosmological perturbations on brane world. In section \ref{2}, we
develop the reconstruction in our model. In Section \ref{3}, we
consider the high energy limit and the reconstruction, considering the
attractor as  the scalar spectral index $n_s(N)$. Here, we  formulate
a general formalism to find the effective potential and
 in Sec. \ref{4},
we also apply our results to a specific example for the spectral
index $n_s(N)$. In section \ref{6}, we formulate the reconstruction
from the tensor to scalar ratio $r(N)$ under a general formalism  and in subsection\ref{7},
we consider as example the attractor $r(N)\propto N^{-2}$. Finally, in section
\ref{conclu} we summarize our findings. We chose units so that
$c=\hbar =1$.

\section{ Braneworld  inflation: basic equations }

In this section we give a brief review of the background equations
and cosmological perturbations on the brane.  We begin with the
action given by
\begin{equation}
S=M_5^3\int\,d^5x\,\sqrt{G}\left(^{(5)}R-2\Lambda_5\right)-\int\,d^4x\,\sqrt{-g}\,L_{matter},\label{S1}
\end{equation}
where the quantity $^{5}R$ corresponds to the  Ricci scalar curvature of the
metric $G_{ab}$ of the five dimensional bulk, $L_{matter}$ describes the matter
confined on the brane, $M_5$ and $\Lambda_5$ are the five-dimensional Planck mass and
cosmological constant, respectively. The relation between  the Planck mass in
four dimensional $m_p$ and $M_5$ and also the relationship  between  the cosmological constants
becomes \cite{2}
$$
m_p=\sqrt{\frac{3}{4\pi}}\,\left(\frac{M_5^2}{\sqrt{\tau}}\right)M_5,\,\,\mbox{and}\,\,\,\,
\Lambda_4=\frac{4\pi}{M_5^4}\left(\Lambda_5+\frac{4\pi}{3M_5^3}\tau^2\right),
$$
respectively. Here, $\Lambda_4$ corresponds to the four dimensional
cosmological constant and the quantity $\tau$ denotes the brane tension.

From the action (\ref{S1}) the authors of  Ref.\cite{3} have shown that the four dimensional Einstein equations induced on
the brane can be written as  (see also Ref.\cite{RM})
\begin{equation}
G_{\mu\nu}=-\Lambda_4\,g_{\mu\nu}+\left(\frac{8\pi}{m_p^2}\right)\,T_{\mu\nu}+
\left(\frac{8\pi}{M_5^2}\right)S_{\mu\nu}-{\mathcal{E}}_{\mu\nu},
\end{equation}
in  which $T_{\mu\nu}$ corresponds to the energy-momentum tensor of the matter,
the quantity $S_{\mu\nu}$ denotes the local correction to standard Einstein eqs.
from the extrinsic curvature and ${\mathcal{E}}_{\mu\nu}$ is the nonlocal effect
correction  due to a free gravitational field which emerges from the projection
of the bulk Weyl tensor. By considering an extended version of  Birkhoff's
theorem, we find  that if the bulk space-time is anti-de Sitter, then the
nonlocal effect corrections ${\mathcal{E}}_{\mu\nu}=0$ \cite{B1} and from the Bianchi identity
($\nabla^\mu G_{\mu\nu}=0$), we have  $\nabla^\mu S_{\mu\nu}=0$\cite{RM}.  On the other hand, assuming that the matter
in the brane
 (the matter is confined in the brane and the gravity
 can be propagated to the extra dimension)
 is describe by a perfect fluid together with a flat FRW metric, then we find that   the  modified
Friedmann equation  becomes \cite{2,3}
\begin{equation}
3H^2=\kappa\,\rho\left[1+\frac{\rho}{2\tau}\right]+\Lambda_4+\frac{\xi}{a^4},
\label{eq1}\end{equation} where the quantity $H=\dot{a}/a$ denotes
the Hubble rate, $a$ corresponds to the scale factor and  $\rho$
denotes  the matter field confined to the brane. Here,   the
constant $\kappa=8\pi/m_p^2$, where $m_p$ is the four-dimensional
Planck mass. The quantity $\xi/a^4$ has a form of dark radiation
and it indicates the influence of the bulk gravitons on the brane,
in which $\xi$ corresponds to an integration constant. As we emphasized before, the
brane tension  $\tau$ is related  with  the four and five
dimensional Planck masses by the relation
$m_p^2=3M_5^6/(4\pi\tau)$ and a constraint on the
value of the brane tension is found from nucleosynthesis given by
$\tau
>$ (1MeV)$^4$ \cite{Cline}.  However, a different constraint for the brane tension from current tests for deviation
from Newton`s law was obtained in Refs.\cite{test1,test2} in which
it is restricted to $\tau\geq $(10
 TeV)$^4$.

In the following, we will consider  that the
 constant $\Lambda_4=0$, and once 
the inflation epoch initiates, the  quantity  $\xi/a^4$ will
rapidly become unimportant, with which   the modified  Friedmann
Eq.(\ref{eq1}) becomes\cite{3}

\begin{equation}
3H^{2}=\kappa\,\rho\left[1+\frac{\rho}{2\tau}\right].  \label{HC}
\end{equation}


In order to describe the matter, we  consider  that  the energy density  $\rho$ corresponds to a standard scalar field $\phi $, where
the energy density $\rho(\phi)$ and the pressure $P(\phi)$ are
defined as
$\rho=\frac{\dot{\phi}^{2}}{2}+V(\phi )$, and   $P=%
\frac{\dot{\phi}^{2}}{2}-V(\phi )$, respectively. Here, the
quantity
 $V(\phi )=V$ denotes the
scalar potential.  We also consider that the scalar field $\phi$
is a homogeneous scalar field i.e., $\phi=\phi(t)$  and also this
field is confined to the brane \cite{2,3}.  In this context, the
dynamics of the scalar field can be written as
\begin{equation}
\dot{\rho}+3H(\rho +P)=0,  \label{key_01}
\end{equation}%
or equivalently
\begin{equation}
\ddot{\phi}+3H\dot{\phi}+V'=0,  \label{ecdf}
\end{equation}%
where $V'=\partial V(\phi)/\partial \phi$. Here  the dots mean
derivatives with respect to the cosmological time.



By assuming the slow roll approximation in which the energy
density $\rho\sim V(\phi)$, then the Eq.(\ref{HC}) reduces
to\cite{2,3}
\begin{equation}
3H^{2}\approx\kappa\,V\left[1+\frac{V}{2\tau}\right], \label{HC2}
\end{equation}
and Eq.(\ref{ecdf}) can be written as
\begin{equation}
3H\dot{\phi}\approx-V'.  \label{ecdf2}
\end{equation}%
Following Ref.\cite{2} we can introduce the slow roll parameters
$\epsilon$ and $\eta$ defined as
\begin{equation}
\epsilon=\frac{1}{2\kappa}\left(\frac{V'}{V}\right)^2\,\frac{(1+V/\tau)}{(1+V/2\tau)^2},\,\,\,\,\,\mbox{and}\,\,\,\,\,\,\,\eta=\frac{1}{\kappa}\frac{V''}{V(1+V/2\tau)}.\label{sp}
\end{equation}
On the other hand, introducing the number of $e$-folding $N$ between
two different values of the time $t$ and $t_e$ gives 
\begin{equation}
N=\int_t^{t_e}\,H\,
dt\simeq\kappa\int_{\phi_e}^{\phi}\,\frac{V}{V'}\left(1+\frac{V}{2\tau}\right)\,d\phi,\label{N1}
\end{equation}
where $t_e$ corresponds to the end of the inflationary stage and here  we
have considered the slow roll approximation.

In the context of the brane world  the power spectrum
${\mathcal{P}_{\mathcal{R}}}$ of the curvature
perturbations assuming  the slow-roll approximation is given by %
\cite{4}.
\begin{equation}
 {\mathcal{P}_{\mathcal{R}}}
 =\left(\frac{H^2}{\dot{\phi}^2}\right)\,\left(\frac{H}{2\pi}\right)^2\simeq\frac{\kappa^3}
 {12\pi^2}\,\frac{V^3}{V'\,^2}\,\left(1+\frac{V}{2\tau}\right)^3.\label{Pet}
\end{equation}
The scalar spectral index $n_{s}$ is defined as $n_{s}-1=\frac{d\ln \,{%
\mathcal{P}_{R}}}{d\ln k}$ and  in terms of the slow roll
parameters $\epsilon$ and $\eta$ can be written as\cite{4}
\begin{equation}
n_s-1=-6\epsilon+2\eta.\label{ns1}
\end{equation}
Here we have used Eqs.(\ref{sp}) and (\ref{Pet}), respectively.

It is well known that  the tensor-perturbation during inflation
would produce gravitational waves. In the braneworld the tensor
 perturbation
is more complicated than the standard expression obtained in  GR, where the
amplitude of the tensor perturbations ${\mathcal{P}}%
_{g}\propto H^2$. Because the braneworld gravitons propagate in
the bulk, the amplitude of the tensor perturbation suffers a
modification \cite{t}, wherewith
\begin{equation}
{\mathcal{P}}%
_{g}=8\kappa \,\left( \frac{H}{2\pi }\right)
^{2}F^{2}(x)\label{PGB},
\end{equation}
where the quantity $x=Hm_{p}\sqrt{3/(4\pi \tau )}$ and the
function $F(x)$ is defined as
\begin{equation}
F(x)=\left[ \sqrt{1+x^{2}}-x^{2}\sinh ^{-1}(1/x)\right] ^{-1/2},
\label{Fx}
\end{equation}
in which  the correction given by  the function $F(x)$, appeared from the normalization of a zero-mode%
\cite{t}. In particular in the limit in which the tension $\tau\gg
V$, the function $F(x)\rightarrow 1$ and then ${\mathcal{P}}%
_{g}\propto H^2$.

An important observational quantity is the tensor to scalar ratio
$r$, defined as
$r=\left(\frac{{\mathcal{P}}_g}{P_{\mathcal{R}}}\right)$. Thus,
combining Eqs.(\ref{Pet}) and (\ref{PGB}), the tensor-scalar
ratio, $r$, is given by
\begin{equation}
r=\left(
\frac{{\mathcal{P}}_{g}}{\mathcal{P}_{\mathcal{R}}}\right)
\simeq\frac{8}{\kappa}\left(\frac{V'}{V}\right)^2\,\left(1+\frac{V}{2\tau}\right)^{-3}\,F^2(V).
\label{Rk1}
\end{equation}%
Here, we have considered that the quantity $x$ can be rewritten in
terms of the effective potential from Eq.(\ref{HC2}).

\section{Reconstruction on brane}\label{2}

In this section we consider  the methodology in order to
reconstruct the background variables, considering the scalar
spectral index in terms of the number of $e$-folds in the framework
of brane-world. As a first part, we rewrite the scalar spectral
index given by Eq.(\ref{ns1}), as a function of the number of
$e$-folds $N$ and its derivatives. In this form, obtaining the index
$n_s=n_s(N)$,
 we should find the potential $V=V(N)$ in terms  of the number of $e$-folding  $N$.
  Subsequently, utilizing the relation given by  Eq.(\ref{N1}), we
should obtain the $e$-folds $N$ as  a function of the scalar field
$\phi$ i.e., $N=N(\phi)$. Finally, considering these relations, we
can reconstruct the effective potential $V(\phi)$ in order to
satisfy a specific  attractor $n_s(N)$.

In this way, we start  by
rewriting the standard slow roll parameters $\epsilon$ and $\eta$
in terms of the number of $e$-folds $N$.  Thus, the derivative of
the scalar potential $V'$ from Eq.(\ref{N1}) can be rewritten as
\begin{equation}
V'=\frac{dV}{d\phi}=V_{,\,N}\frac{dN}{d\phi},\,\,\,\,\,\mbox{in
which}\,\,\,\,V'\,^2=\kappa V\,\left(1+\frac{V}{2\tau}\right)\,V_{,\,N},\label{dV}
\end{equation}
and this suggests that $V_{,\,N}$ is a positive quantity.
 In the following, we will consider that the
notation $V_{,\,N}$ corresponds to $dV/dN$, $V_{,\,NN}$ denotes $d^2V/dN^2$, etc.

Analogously, we can rewrite $V''$ as
$$
V''=\frac{\kappa}{2V_{,\,N}}\,\left[V_{,\,N}^2\,\left(1+\frac{V}{\tau}\right)+V\left(1+
\frac{V}{2\tau}\right)V_{,\,NN}\right].
$$

In this form, the slow roll parameter $\epsilon$ can be rewritten  as
\begin{equation}
  \epsilon=\frac{1}{2}\,\frac{\left(1+\frac{V}{\tau}\right)}{V\left(1+\frac{V}{2\tau}\right)}\,\,V_{,\,N},
\end{equation}
and the parameter $\eta$ as
\begin{equation}
\eta=\frac{1}{2V}\,\frac{\left(1+\frac{V}{\tau}\right)}{\left(1+\frac{V}{2\tau}\right)}\,V_{,\,N}+
\frac{V_{,\,NN}}{2V_{,\,N}}\,\,,
\end{equation}
respectively. Here, we have considered that $V'>0$.

Also, from Eqs.(\ref{N1}) and (\ref{dV}) we can rewritten $dN/d\phi$ as
\begin{equation}
\frac{dN}{d\phi}=\sqrt{\frac{\kappa
V}{V_{,\,N}}}\,\sqrt{\left(1+\frac{V}{2\tau}\right)}\,.\label{Nf}
\end{equation}

In this way, by using  Eq.(\ref{ns1}) we find that the scalar
spectral index can be rewritten as
\begin{equation}
  n_s-1=-2\frac{\left(1+\frac{V}{\tau}\right)}{V\left(1+\frac{V}{2\tau}\right)}\,V_{,\,N}+
  \frac{V_{,\,NN}}{V_{,\,N}},
\end{equation}
or equivalently
\begin{equation}
n_s-1=-2\left[\ln\left(V\left[1+\frac{V}{2\tau}\right]\right)\right]_{,\,N}+[\ln\,V_{,\,N}]_{,\,N}=
\left[\ln\left(\frac{V_{,\,N}}{V^2\left(1+\frac{V}{2\tau}\right)^2}\right)\right]_{,\,N}.\label{R1}
\end{equation}
We also note that in the limit in which $\tau\rightarrow \infty$,
Eq.(\ref{R1}) reduces to GR, in which $n_s-1=\left(\ln
\frac{V_{,\,N}}{V^2}\right)_{,\,N}$, see Ref.\cite{Chiba:2015zpa}.

From Eq.(\ref{R1}) we have
\begin{equation}
\frac{V_{,\,N}}{V^2(1+V/2\tau)^2}=e^{\int\,{(n_s-1)}dN}.\label{dVN}
\end{equation}
This equation gives us the effective potential $V(N)$ for a
specific attractor $n_s(N)$. Thus, integrating we have
\begin{equation}
  \frac{1}{\tau}\,\ln\left(\frac{1+V/2\tau}{V/2\tau}\right)-\frac{(1+V/\tau)}{V(1+V/2\tau)}=
 \int \left[e^{\int\,{(n_s-1)}dN}\right]\,dN\,.
\end{equation}
However, this equation results in a transcendental equation for the scalar
potential $V$ and this result does not permit one to obtain the relation $V=V(N)$.

We also note that by combining  Eqs.(\ref{Nf}) and (\ref{dVN}), we
obtain that the relation between  the number of $e$-folds $N$ and
the scalar field $\phi$  can be written as
\begin{equation}
\left[\sqrt{V\,\left(1+\frac{V}{2\tau}\right)}\,e^{\int\,{\frac{(n_s-1)}{2}dN}}\right]\,dN=d\phi.
\end{equation}

On the other hand, from Eq.(\ref{Rk1}) the tensor-scalar ratio,
$r$ can be rewritten as
\begin{equation}
r(N)
\simeq\left(\frac{4}{\tau}\right)\,V_{,\,N}\,\left(1+\frac{V}{2\tau}\right)^{-3}\,F^2(V).
\label{Rk2} \end{equation}

In the following we will consider the high energy limit in which
$\rho\simeq V\gg\tau$,  in order to obtain an  analytical solution in
the reconstruction of the scalar potential in terms of the scalar
field $V(\phi)$.

\section{High energy:Reconstruction from the attractor
$n_s(N)$}\label{3}

In this section we consider  the high energy limit ($V\gg\tau$) in
order to reconstruct the scalar potential, considering as an 
attractor the scalar spectral index in terms of the number of
$e$-folds i.e., $n_s=n_s(N)$.  In this limit, the derivatives $V'$
and $V''$ can be rewritten as
\begin{equation}
  V'\,^2=\frac{\kappa}{2\tau}\,V^2\,V_{,\,N},\,\,\,\,\mbox{and}\,\,\,\,V''=\frac{\kappa}{2\tau}\,V\,
  \left[V_{,\,N}+\frac{V\,V_{,\,NN}}{2V_{,\,N}}\right].
\end{equation}
In this way, the relation between the number $N$ and the scalar
field $\phi$ in this limit becomes
\begin{equation}
  \frac{dN}{d\phi}=\frac{\kappa}{2\tau}\,\left(\frac{V^2}{V'}\right)=
  \left(\frac{\kappa}{2\tau}\right)^{1/2}\,\frac{V}{\sqrt{V_{,\,N}}}.\label{NF2}
\end{equation}
From Eq.(\ref{ns1}) we find that the scalar spectral index $n_s$ results in 
\begin{equation}
n_s-1=\frac{4\tau}{\kappa}\left[V''-\frac{3V'\,^2}{V}\right]\,\frac{1}{V^2}=-4\frac{V_{,\,N}}{V}
+\frac{V_{,\,NN}}{V_{,\,N}},
\end{equation}
or equivalently
\begin{equation}
n_s-1=-4[\ln\,V]_{,\,N}+[\ln
V_{,\,N}]_{,\,N}=\left[\ln\left(\frac{V_{,\,N}}{V^4}\right)\right]_{,\,N}.\label{nN2}
\end{equation}
We note that the relation between the scalar potential and the scalar spectral
index given by Eq.(\ref{nN2})
 becomes independent  of the brane tension $\tau$ in the high energy limit.

From Eq.(\ref{nN2}), the scalar potential in terms of the number of $e$-foldings can be
written as
\begin{equation}
V=V(N)=\left[-3\int\left(e^{\int(n_s-1)dN}\right)\,dN\right]^{-1/3},\label{VN2}
\end{equation}
where $\int\left(e^{\int(n_s-1)dN}\right)\,dN<0$, in order to make
certain that the potential $V(N)>0$.

Now, by combining  Eqs.(\ref{NF2}) and (\ref{nN2}), we find that
the relation between $N$ and $\phi$ is given by the general expression
\begin{equation}
\left[V\,e^{\int \frac{(n_s-1)}{2}dN}\right]\,dN=\left(\frac{\kappa}{2\tau}\right)^{1/2}\,d\phi,\label{ff}
\end{equation}
where $V$ is given by Eq.(\ref{VN2}).

In this form, Eqs.(\ref{VN2}) and (\ref{ff}) are the fundamental relations in order to
build the scalar potential $V(\phi)$ for an attractor point $n_s(N)$, in the framework of the high energy limit
 in brane world inflation.

On the other hand, in the high energy limit in which $V\gg \tau$,
the function $F^2(x)$ given by  Eq.(\ref{Rk1}) becomes
$F^2(x)\approx\frac{3}{2}x=\frac{3}{2}\frac{V}{\tau}$. In this
form, in the high energy limit the tensor to scalar ratio $r$
becomes
\begin{equation}
r\simeq\,48\tau\,\left(\frac{V_{,\,N}}{V^2}\right).\label{r11}
\end{equation}
Here we have considered Eq.(\ref{Rk2}).

\subsection{An example of $n_s=n_s(N)$.}\label{4}
In order to develop the reconstruction of the scalar potential
$V(\phi)$ in the brane world inflation, we consider the famous
attractor $n_s(N)$ given by
\begin{equation}
  n_s(N)=n_s=1-\frac{2}{N},\label{nsN}
\end{equation}
as example.

From the attractor (\ref{nsN}), we find that considering
Eq.(\ref{nN2}) we have $\frac{V_{,\,N}}{V^4}=\alpha/N^2$, in which
$\alpha$ corresponds to a constant of integration (with units of $m_p^{-12}$) and
 since $V_{,\,N}>0$, then the constant of integration  $\alpha>0$. In this form, the effective
potential as function of the number of $e$-foldings $N$ from
Eq.(\ref{VN2}) becomes
\begin{equation}
V(N)=3^{-1/3}\,\left[\frac{\alpha}{N}+\beta\right]^{-1/3},\label{poth}
\end{equation}
where $\beta$ denotes a new constant of integration. Here, the new
constant of integration $\beta$ with units of $m_p^{-12}$, can be
considered  $\beta=0$ or $\beta\neq 0$.

In the high energy limit, we find  that the power spectrum  ${\mathcal{P}_{\mathcal{R}}}$  given by Eq.(\ref{Pet})
 can be rewritten as
\begin{equation}
{\mathcal{P}_{\mathcal{R}}}\simeq\frac{1}{12\pi^2}\,\left(\frac{\kappa V^2}{2\tau}\right)^3\,\frac{1}{V'\,^2}=
\frac{\kappa^2}{48\pi^2\tau^2}\,\left(\frac{V^4}{V_{,\,N}}\right)=\frac{\kappa^2}{48\pi^2\tau^2}\,
\left(\frac{N^2}{\alpha}\right).\label{P3}
\end{equation}
Note that this result does not depend of the constant of integration
$\beta$. From Eq.(\ref{P3}), it is possible to write the
constant of integration $\alpha$ in terms of the number $N$,
${\mathcal{P}_{\mathcal{R}}}$  and the tension $\tau$ as
\begin{equation}
\alpha=\frac{\kappa^2}{48\pi^2\tau^2}\,
\left(\frac{N^2}{{\mathcal{P}_{\mathcal{R}}}}\right).\label{aa}
\end{equation}
In particular by considering $N=60$ and ${\mathcal{P}_{\mathcal{R}}}=2.2\times
10^{-9}$, we obtain that the  constant of integration $\alpha\simeq 3\times 10^9
(\kappa/\tau)^2$.

On the other hand, from Eq.(\ref{r11}) the tensor to scalar ratio
can be rewritten as
\begin{equation}
r\simeq\,48\tau\,\left(\frac{V_{,\,N}}{V^2}\right)=48\tau\alpha\,\left(\frac{V^2}{N^2}\right)=48\tau\alpha\,\left(\frac{3^{-2/3}}{N^2(\alpha/N+\beta)^{2/3}}\right).
\end{equation}
Note that considering the attractor $n_s$ given by Eq.(\ref{nsN}),
we can find a relation between the tensor to scalar ratio $r$ with
the scalar spectral index or the consistency relation becomes
\begin{equation}
r(n_s)\simeq
\left(\frac{12\alpha\tau}{3^{2/3}}\right)\,\left[\frac{\alpha(1-n_s)}{2}+
\beta\right]^{-2/3}\,(1-n_s)^2.\label{37}
\end{equation}

In the following, we will analyze the cases separately  in which the constant of  integration $\beta$
takes the values $\beta=0$ and $\beta\neq 0$, in order to reconstruct the effective
potential $V(\phi)$.

For the case $\beta=0$, we obtain that the relation between the number of
$e$-foldings $N$ and scalar  field $\phi$ considering  Eqs.(\ref{ff}), (\ref{nsN}) and (\ref{poth})
becomes
\begin{equation}
  N(\phi)=N=\frac{1}{3^2\alpha^{1/2}}\,\left(\frac{\kappa}{2\tau}\right)^{3/2}\,\,(\phi-\phi_0)^3,\label{N12}
\end{equation}
where $\phi_0$ corresponds to a constant of  integration. In this way, in the high energy limit  we find that
the reconstruction  of the effective potential as a function of the scalar field for the case $\beta=0$
and assuming the attractor $n_s-1=-2/N$
 is
given by
\begin{equation}
V(\phi)=V_0\,\,(\phi-\phi_0),\,\,\,\mbox{where}\,\,\,\,\,\,\,V_0=\left(\frac{\kappa}{18\tau\alpha}\right)^{1/2}.\label{P12}
\end{equation}
Also, we note that for the case $\beta=0$, the consistency relation $r=r(n_s)$
has a dependence  $r(n_s)\propto (1-n_s)^{4/3}$. In particular, by considering $n_s=0.964$, $N=56$
and
${\mathcal{P}_{\mathcal{R}}}=2.2\times
10^{-9}$, we find an upper bound for the brane tension  given by $\tau<
10^{-13}m_p^4$,
 from the condition $r<0.07$.  For this bound on $\tau$, we have used
 Eq.(\ref{37}). Now, from Eq.(\ref{aa}) and considering  $N=60$ and ${\mathcal{P}_{\mathcal{R}}}=2.2\times
10^{-9}$, together with the upper limit on $\tau$, we obtain a
lower limit for the constant $\alpha$ given by $\alpha>1.9\times
10^{38}m_p^{-12}$.

\begin{figure}[th]
{{\vspace{0.0 cm}\includegraphics[width=4.5in,angle=0,clip=true]{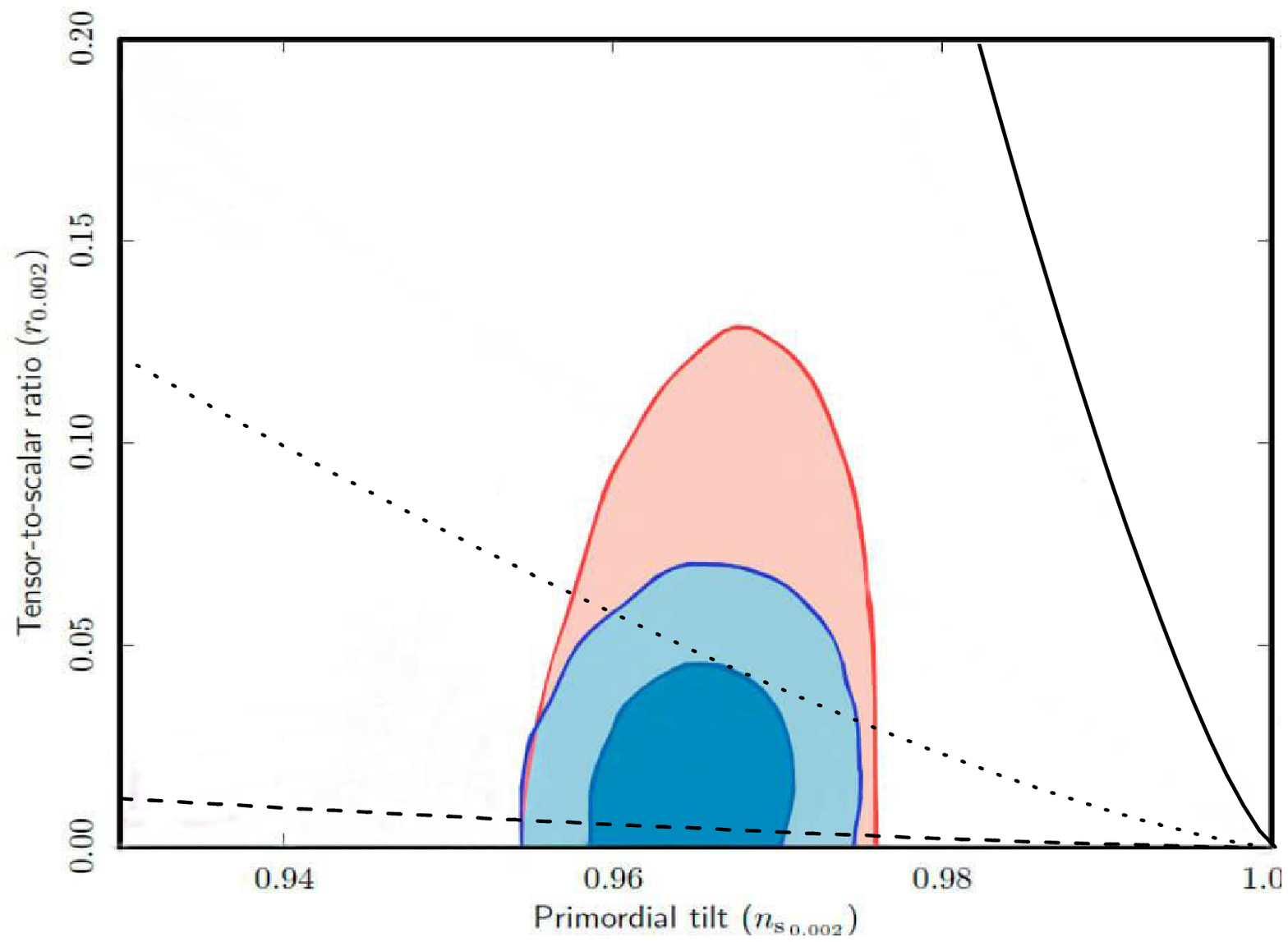}}}
{{\vspace{-1.5 cm}{\includegraphics[width=4.5in,angle=0,clip=true]{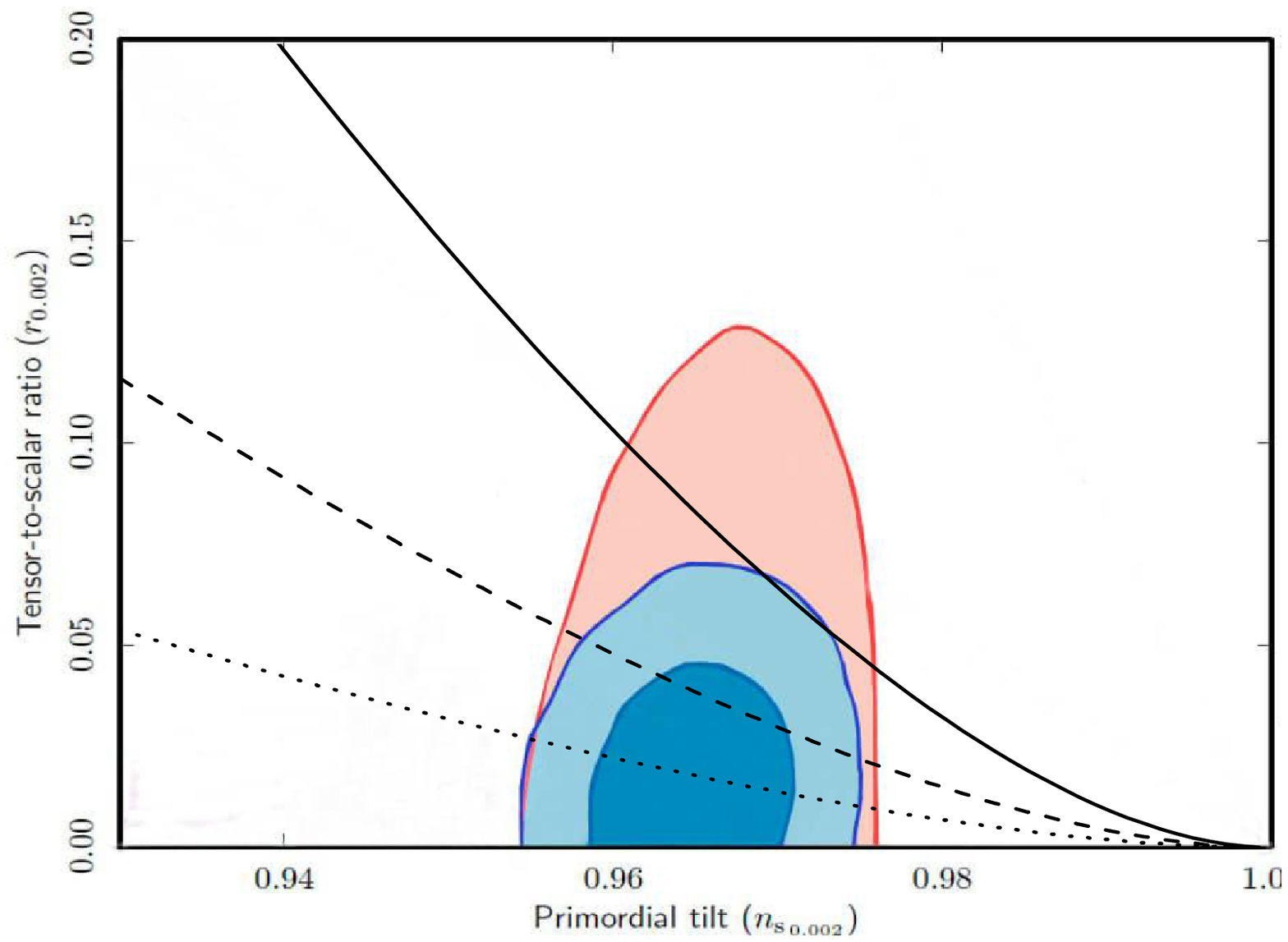}}}}
{\vspace{-1.0 cm}\caption{  The upper and lower panels show the tensor-to-scalar ratio $r$ as a
function of the scalar spectral index $n_s$,  for
three different values of the brane tension $\tau$.  In both panels we have
considered the two-marginalized constraints jointly as  68$\%$ and 95$\%$ C.L. at $k=0.002$ Mpc$^{-1}$ from the Planck 2018 results
\cite{Planck2018}. Also,
in both panels the solid,
dashed and dotted lines correspond to the values of brane tension
$\tau/m_p^4=10^{-12},10^{-13}$ and $10^{-14}$, respectively. In the upper panel we show
the consistency relation  for  the specific
case in which the constant $\beta=0$ and we have used
$\alpha=10^{38}m_p^{-12}$. In the lower panel we show the case in which $\beta\neq 0$
and we have considered
$\beta=\alpha/60$ and $\alpha=3\times 10^{9}(\kappa/\tau)^2$, respectively.
 \label{fig1}}}
\end{figure}

On the other hand, in the reconstruction for the situation in which the constant of  integration $\beta\neq 0$, we find that
  considering  Eq.(\ref{ff})   the relation between $dN$ and $d\phi$ can be written as
\begin{equation}
\frac{dN}{[\alpha N^2+\beta
N^3]^{1/3}}=\frac{dN}{\beta^{1/3}[\alpha_0 N^2+
N^3]^{1/3}}=C_1 d\phi,\,\,\,\,\;\:\;\mbox{where}\,\,\,\,\,\;\;\;
C_1=3^{1/3}\left(\frac{\kappa}{2\alpha\tau}\right)^{1/2},\label{NF6}
\end{equation}
and the quantity $\alpha_0=\alpha/\beta$. In the following, we will consider  for simplicity the case in which
 the  constant of  integration $\beta>0$ i.e, $\alpha_0>0$.
We also note that the integration of  Eq.(\ref{NF6}) does not permit one to
obtain an analytical solution for the number of $e$-folds as a function of the scalar field i.e.,
$N=N(\phi)$. In this sense,
the solution of Eq.(\ref{NF6}) can be written as
$$
  \sqrt{3}\arctan\left[\left(1+2\left[1+\frac{\alpha_0}{N}\right]^{-1/3}\right)/\sqrt{3}\right]+\frac{1}{2}\ln
  \left[1+\left(1+\frac{\alpha_0}{N}\right)^{-2/3}+\left(1+\frac{\alpha_0}{N}\right)^{-1/3}\right]
$$
\begin{equation}
  -\ln\left[1-\left(1+\frac{\alpha_0}{N}\right)^{-1/3}\right]=\beta^{1/3}\,C_1\,(\phi-\phi_0),
  \label{sol3}
\end{equation}
where $\phi_0$ denotes a constant of  integration.

\begin{figure}[th]
{{\hspace{0cm}\includegraphics[width=3.2in,angle=0,clip=true]{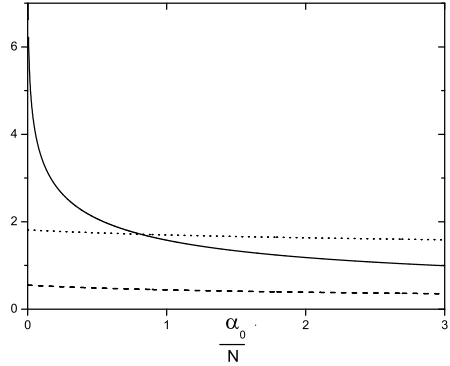}}}
{\vspace{-0.1 cm}\caption{  Evolution of the three terms on the right given
by Eq.(\ref{sol3}) versus the dimensionless quantity
$\frac{\alpha}{\beta\,N}=\frac{\alpha_0}{N}$. Here, the  dotted, dashed, and solid lines
denote the first, second and third terms of Eq.(\ref{sol3}), respectively.
 \label{fig2}}}
\end{figure}

Numerically, we note that in the limit in which
$\alpha/\beta=\alpha_0<N$, the first two terms of Eq.(\ref{sol3})
are approximately constants and the dominant term corresponds to (see Fig.\ref{fig2} )
\begin{equation}
  -\ln\left[1-\left(1+\frac{\alpha_0}{N}\right)^{-1/3}\right]\approx\beta^{1/3}\,C_1\,(\phi-\phi_0).\label{app2}
\end{equation}
Thus, we find that the reconstruction of the effective potential $V(\phi)$
considering the specific case in which $\alpha_0<N$ is given by
\begin{equation}
  V(\phi)\approx\frac{1}{(3\,\beta)^{1/3}}\,\,\left[1-\exp(-\beta^{1/3}\,C_1\,[\phi-\phi_0])\right].\label{pp8}
\end{equation}
Here, we have combined  Eqs.(\ref{poth}) and (\ref{app2}).
Curiously, we observe  that this effective potential is similar to
that obtained in the Starobinsky model \cite{Staro} in which
$\beta^{1/3}\,C_1=\sqrt{2/3}\,m_p^{-1}$ i.e.,
$\beta^{1/3}\,C_1\,m_p \approx\mathcal{O}(1)$. Also, in the limit
$\beta^{1/3}\,C_1\,[\phi-\phi_0]\gg 1$, the effective potential
corresponds to a  constant potential i.e., a solution of de
Sitter.

For the inverse case in which $\alpha_0>N$, we note that
 the first term of Eq.(\ref{sol3}) dominates with
which (see Fig.\ref{fig2} )
\begin{equation}
  \sqrt{3}\arctan\left[\left(1+
  2\left[1+\frac{\alpha_0}{N}\right]^{-1/3}\right)/\sqrt{3}\right]\approx\beta^{1/3}\,C_1\,(\phi-\phi_0).
\end{equation}

In this form, we obtain that the reconstruction in the limit in which $\alpha_0>N$
becomes
\begin{equation}
  V(\phi)\approx\frac{1}{2(3\beta)^{1/3}}\left[\sqrt{3}\tan(\beta^{1/3}C_1(\phi-\phi_0)/\sqrt{3})-1\right].\label{pp6}
\end{equation}
Here the range for the scalar field is given by
$\frac{\sqrt{3}\,\pi}{6\beta^{1/3}C_1}+\phi_0\lesssim \phi
\lesssim \frac{\sqrt{3}\,\pi}{2\beta^{1/3}C_1}+\phi_0$.

In Fig.\ref{fig1} we show the ratio $r$ versus the  spectral index $n_s$, for three different values
of the brane tension $\tau$. In both panels we consider the two-marginalized constraints for
the consistency relation $r=r(n_s)$ (at 68$\%$ and 95$\%$ CL at  $k=0.002$ Mpc$^{-1}$ ) from the new
Planck data \cite{Planck2018}. In the upper panel we consider the special case
in which the constant of integration $\beta=0$, where the consistency relation is given by  Eq.(\ref{37}). Here, we take the value
$\alpha=10^{38}m_p^{-12}$.  In the lower panel we take into account the case in
which $\beta\neq 0$ and for the relation $r=r(n_s)$ we have used Eq.(\ref{37}). In this case
 we have considered the specific value of $\beta$ at $N=60$ (point limit $\alpha_0/N=1$ or $\beta=\alpha/N$)
 wherewith
$\beta=\alpha/60$ and $\alpha=3\times 10^{9}(\kappa/\tau)^2$, respectively.
Also, in both panels the solid,
dashed and dotted lines correspond to the values of brane tension
$\tau/m_p^4=10^{-12},10^{-13}$ and $10^{-14}$, respectively. In particular for  the
case $\beta=0$ we find that the brane tension has an upper limit given by $\tau<10^{-13}
m_p$, as can be seen of the upper panel of Fig.\ref{fig1}. For the case in which
$\beta\neq 0$ we find that in the particular case in which $\beta=\alpha/60$, the
value of the brane tension $\tau<10^{-12}m_p^4$ is well corroborated by Planck
2018 results, see lower panel of Fig.\ref{fig1}.
This suggests that the value of the constant of integration $\beta$ modifies the
the upper bound on the brane tension. We note that in the case in which the
constant $\beta>\alpha/60$ the upper limit on the brane tension increases and  in the
opposite case ($\beta<\alpha/60$
) the upper limit on $\tau$ decreases.

In Fig.\ref{fig2} we show the behavior of the three terms on the right of
 Eq.(\ref{sol3}) versus the dimensionless quantity
$\frac{\alpha}{\beta\,N}=\frac{\alpha_0}{N}$. We note that for the limit in
which $\alpha_0<N$ dominates the third term of  Eq.(\ref{sol3}), see solid line of Fig.\ref{fig2}. However, for
the case in which $\alpha_0>N$ the dominant term  corresponds to the first
expression of
 Eq.(\ref{sol3})  given by  dotted line in Fig.\ref{fig2}.

In order to clarify our above results, we can study some specific
limits for the ratio $\alpha/(\beta N)=\alpha_0/N$ in which
$\alpha_0/N\ll 1$ and $\alpha_0/N\gg 1$. As a first approximation we
consider the case in which $\alpha_0/ N \ll 1$ or $\alpha_0\ll N$.
For this limit we find from Eq.(\ref{NF6}) that the relation
$N=N(\phi)$ is given by
\begin{equation}
  N(\phi)=\exp[\beta^{1/3}\,C_1\,(\phi-\phi_0)],
\end{equation}
where $\phi_0$ denotes a constant of integration. Thus, considering
the limit  $\alpha/(\beta  )=\alpha_0\ll N$ we obtain that the
effective potential $V(\phi)$  given by Eq.(\ref{poth}) becomes a
constant and equal to $V(\phi)=(3\beta)^{-1/3}$. In fact, this
result indicates an accelerated expansion de Sitter or de Sitter
inflation, since in the high energy limit and considering  the
slow-roll approximation,  we have $H\propto V=$constant. Note that
this constant potential coincides with the potential given by
Eq.(\ref{pp8}) when $\beta^{1/3}\,C_1\,[\phi-\phi_0]\gg 1$. We also observe that
for the consistency relation $r=r(n_s)$, we get $r=48\tau
\alpha/[(3\beta)^{2/3}N^2]\propto (1-n_s)^2$ (see Eq.(\ref{37})).

For the  case in which  $\alpha/(\beta N )\gg 1$ or $\alpha_0\gg N$, we find from Eq.(\ref{NF6}) that the relation $N=N(\phi)$
coincides with the case $\beta=0$ i.e., Eq.(\ref{N12}) and then
the effective potential $V(\phi)$ changes linearly with the scalar
field according to Eq.(\ref{P12}) in which $V(\phi)\propto \phi$.
This effective potential agrees with the potential given by
Eq.(\ref{pp6}) assuming that the argument
$\beta^{1/3}C_1(\phi-\phi_0)/\sqrt{3}<1.$

\section{High energy:Reconstruction from the attractor
$r(N)$}\label{6}

In this section we consider the hight energy limit, in order to reconstruct the
effective potential $V(\phi)$, but from a different point of view. In order to
reconstruct the scalar potential, we consider as an attractor the tensor to scalar
ratio in terms of the number of $e-$foldings $N$ i.e., $r=r(N)$. In this sense,
considering Eq.(\ref{r11}) we obtain that the potential effective $V(N)$ can be written as
\begin{equation}
 V= V(N)=-48\tau \,\left[\int\,r\,dN\right]^{-1}.\label{VN3}
\end{equation}

Now from Eq.(\ref{NF2}) we find that the relation between the number $N$ and the scalar field $\phi$
is given by
\begin{equation}
  r^{1/2}\,\frac{dN}{d\phi}=(24\kappa)^{1/2}.\label{Nrb}
\end{equation}
Here, we have considered Eq.(\ref{r11}).

In this context, we can obtain the scalar spectral index $n_s$ as a function of
the number of $e$-folds $N$, combining the expressions given  by Eqs.(\ref{nN2}) and (\ref{VN3}) for
 a specific  attractor $r=r(N)$. Thus, the scalar spectral index can be
 rewritten as
\begin{equation}
  n_s-1=\,\left[\ln\left(\frac{r}{48\tau
  V^2}\right)\right]_{,\,N}.\label{ns4}
\end{equation}
Here, the potential $V$ is given by Eq.(\ref{VN3}).

\subsection{An example of $r=r(N)$.}\label{7}

In order to develop the reconstruction of the scalar potential
$V(\phi)$ in the brane world inflation, we consider that the attractor
for the tensor to scalar ratio as a function of the number of $e$-
folds  $r(N)$ is given by
\begin{equation}
  r(N)=\frac{\alpha_1}{N^2},\label{rN3}
\end{equation}
where $\alpha_1>0$ corresponds to a constant (dimensionless).  For
this attractor the cases in which $\alpha_1=12$, was analyzed in
Ref.\cite{T}, and the specific value $\alpha_1=8$, was obtained in
Ref. \cite{Chiba:2015zpa}.

In
particular considering $N=60$ and $r<0.07$, we find that the value
of the constant  $\alpha_1<252$.

By combining Eqs.(\ref{VN3}) and (\ref{rN3}) we obtain that the
scalar  potential in terms of the number of $e$-foldings becomes
\begin{equation}
V(N)=\frac{1}{\alpha_2/N+\beta_1},\,\,\,\,\,\,\,\mbox{where}\,\,\,\,\,\,\alpha_2=\frac{\alpha_1}{48\tau}.\label{Pot5}
\end{equation}
Here the quantity $\beta_1$ corresponds to a constant of integration
with units of $m_p^{-4}$.

In order to obtain the relation between the number $N$ and the
scalar field $\phi$, we consider Eq.(\ref{Nrb}) together with the
attractor given by  Eq.(\ref{rN3}) obtaining
\begin{equation}
N=\exp\left[\sqrt{\frac{24\kappa}{\alpha_1}}\,(\phi-\phi_0)\right],
\end{equation}
where $\phi_0$ denotes a new constant of integration. Thus, the
reconstruction of the scalar potential in terms of the scalar
field can be written as
\begin{equation}
V(\phi)=\left(\alpha_2\,\exp\left[-\sqrt{\frac{24\kappa}{\alpha_1}}
\,(\phi-\phi_0)\right]+\beta_1\right)^{-1}.\label{pote5}
\end{equation}

In particular assuming that $\beta_1>0$ and $\alpha_2/\beta_1\gg
N$, the effective potential has the behavior of an exponential
potential i.e., $V(\phi)\propto
e^{(\sqrt{24\kappa/\alpha_1}\,)\,\phi}$ (recall the we have considered that $V'>0$). In the inverse case in
which $N\gg \alpha_2/\beta_1$, the scalar potential corresponds to
a constant potential $V(\phi)=$ constant.

In the context of the cosmological perturbations, we find in the high energy limit the power spectrum
becomes
\begin{equation}
{\mathcal{P}_{\mathcal{R}}}\simeq\frac{1}{12\pi^2}\,\left(\frac{\kappa
V^2}{2\tau}\right)^3\,\frac{1}{V'\,^2}=
\frac{\kappa^2}{48\pi^2\tau^2}\,\left(\frac{V^4}{V_{,\,N}}\right)=\frac{\kappa^2}{48\pi^2\tau^2}\,\frac{N^2}{\alpha_2}\,\left(\frac{\alpha_2}{N}+\beta_1\right)^{-2}.
\end{equation}
Here we have used Eqs.(\ref{Pet}) and (\ref{Pot5}), respectively.
Thus, we can write the  constant $\beta_1$  in terms of
the scalar spectrum ${\mathcal{P}_{\mathcal{R}}}$, the number of
$e$-folds $N$ and the constant $\alpha_2$ as
\begin{equation}
\beta_1=\sqrt{\frac{1}{3\,\alpha_2\,{\mathcal{P}_{\mathcal{R}}}}}\,\left(\frac{\kappa\,N}{4\pi\,\tau}\right)-\frac{\alpha_2}{N}
=\sqrt{\frac{\alpha_2}{3\,\,{\mathcal{P}_{\mathcal{R}}}}}\,\left(\frac{12\kappa\,N}{\pi\,\alpha_1}\right)-\frac{\alpha_2}{N}.\label{a10}
\end{equation}

On the other hand, from Eq.(\ref{ns4}) we find that the relation between the scalar
index $n_s$ and the number of $e$-foldings is given by
\begin{equation}
n_s-1=\frac{2}{N}\left[\frac{\beta_1}{\alpha_2/N+\beta_1}-2\right].\label{nsN6}
\end{equation}

Note that in the specific case in which $N\gg \alpha_2/\beta_1$,
the scalar spectral index $n_s$ gives the famous attractor
$n_s-1=-2/N$.

Now, from Eq.(\ref{nsN6}) we can find the constant $\beta_1$ in terms
of  $n_s$, $N$ and $\alpha_2$ as
\begin{equation}
\beta_1=\frac{[N(n_s-1)+4]}{N[(1-n_s)N-2]}\,\alpha_2.\label{a12}
\end{equation}
Note that for the values  $n_s=0.964$ and $N=60$, we have that the
ratio $\alpha_2/\beta_1\sim 5$. This suggests that the limit $\alpha_2/\beta\gg N$
is not satisfied for large $N$, then the exponential potential $V(\phi)\propto e^\phi$
does not work in the  braneworld. This analysis for the exponential potential 
 in the framework of a brane
 coincides with that obtained in
Ref.\cite{Tsujikawa:2003zd}.  Thus, the reconstruction of the effective potential $V(\phi)$
is given by Eq.(\ref{pote5}) for large $N$ and an appropriate limit corresponds to $N\gg \alpha_2/\beta_1$,
where the behavior of the  scalar potential becomes  constant.

In this form, combining Eqs.(\ref{a10}) and (\ref{a12}) we find
that the tension $\tau$ as function of the observables $n_s$ and
${\mathcal{P}_{\mathcal{R}}}$ together with the number of
$e$-foldings $N$ and $\alpha_1$ becomes
\begin{equation}
\tau=\left(\frac{{\mathcal{P}_{\mathcal{R}}}\;\pi^2\,\alpha_1^3}
{4\,\times12^3\,\kappa^2\,N^4}\right)\,\,\left[\frac{[N(n_s-1)+4]}{[(1-n_s)N-2]}+1\right]^2.
\end{equation}
Here we have used that $\alpha_2=\alpha_1/(48\tau)$.

In particular assuming that the spectral  index $n_s=0.964$,
 the spectrum ${\mathcal{P}_{\mathcal{R}}}\simeq 2.2\times 10^{-9}$ and $N=60$, we
 obtain that the constraint on
 the brane tension $\tau$ is given by
\begin{equation}
  \tau\simeq\, 6\times 10^{-20}\,\alpha_1^3\,\,m_p^4.\label{rel}
\end{equation}
Note that Eq.(\ref{rel}) gives a relation between the brane tension and the
parameter $\alpha_1$. Now,
by assuming that $\alpha_1<252$ in order to obtain $r<0.07$ at $N=60$, we find
that the upper bound for the brane tension becomes
$$
\tau<9.6\times 10^{-13} \,m_p^4\simeq 10^{-12}\,m_p^4.
$$

On the other hand, from Eq.(\ref{nsN6}) we find that the relation
between the scalar index and the tensor to scalar ratio, can be
written as
\begin{equation}
n_s-1=-\frac{2\,r^{1/2}}{\alpha_1^{1/2}}\,
\left[\frac{2\alpha_2+\beta_1\sqrt{\alpha_1/r}}{\alpha_2+\beta_1\sqrt{\alpha_1/r}}\right].\label{rr4}
\end{equation}
Here we have used the attractor given by Eq.(\ref{rN3}).

\begin{figure}[th]
{{\hspace{0cm}\includegraphics[width=4.5in,angle=0,clip=true]{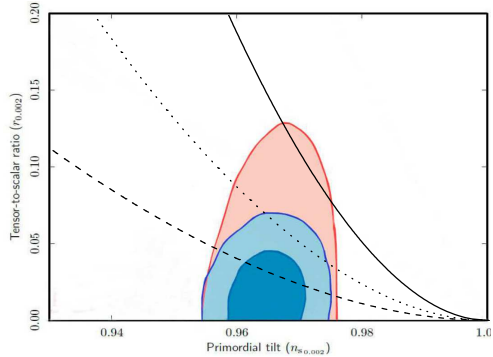}}}
{\vspace{-1.0 cm}\caption{  As before, we show the tensor-to-scalar ratio $r$ as a
function of the scalar spectral index $n_s$ from Planck 2018 results\cite{Planck2018}  for
three different values of the brane tension $\tau$ but assuming the attractor 
$r(N)\propto N^{-2}$ as the starting point.
Solid,  dotted and
dashed   lines correspond to the values of brane tension
$\tau/m_p^4=10^{-11},10^{-12}$ and $10^{-13}$, respectively.
 \label{fig3}}}
\end{figure}

In Fig.\ref{fig3} we show the tensor to scalar ratio versus the scalar spectral
index for three different values of the brane tension considering the attractor $r(N)=\alpha_1\,
N^{-2}$. Here we have used Eq.(\ref{rr4}) and the solid,  dotted and
dashed   lines correspond to the values of brane tension
$\tau/m_p^4=10^{-11},10^{-12}$ and $10^{-13}$, respectively. From this plot we check that the
upper limit for the brane tension given by $\tau<10^{-12}m_p^4$ is well  corroborated
from Planck data.

\section{Conclusions \label{conclu}}

In this article we have analyzed the reconstruction of the background
in the context of braneworld inflation.
Considering  a general formalism of reconstruction, we have obtained  an
expression for the effective potential under
  the slow roll
approximation. In order to obtain analytical solutions in the reconstruction on the brane,
 we
have considered the high energy  limit in which the energy density
$\rho\simeq V\gg \tau$. In this analysis for the reconstruction of
the background, we have considered the parametrization of the
scalar spectral index or the tensor to scalar ratio as function of
the number of $e$-foldings $N$.
 In
this general description  we have found from the cosmological
parameter $n_s(N)$ or the parameter $r(N)$, integrable solutions
 for the effective potential depending on the cosmological attractor $n_s(N)$ or $r(N)$.

  For the reconstruction from the attractor associated with scalar spectral index
  $n_s(N)$, we have assumed the famous attractor $n_s=1-2/N$ as an example.
  From this attractor, we have obtained that the consistency relation $r=r(n_s)$
  is given by Eq.(\ref{37}) and from the power spectrum we have found that the
  integration constant $\alpha$ depends on the brane tensor, see Eq.(\ref{aa}).
  On the other hand,
  depending on the   value of the second  constant of integration $\beta$,
  we have found different results for the reconstruction of the effective potential $V(\phi)$.
  In particular for the specific case in which the constant $\beta=0$, we have
  obtained that the reconstruction of the effective potential corresponds to a potential  $V(\phi)\propto\,\phi$. Also, assuming that
  the observational constraint  on the tensor to scalar ratio $r<0.07$, we have found an upper
  limit for the brane tension given by $\tau< 10^{-13}m_p^4$, wherewith the brane model
  is well supported by the Planck data, see upper panel of Fig.\ref{fig1}. In
  this same context,
  for the case in
  which the  constant of integration $\beta\neq 0$, we have found a
  transcendental equation for the number of $e$-folds as a  function of the
  scalar field  $N=N(\phi)$ and the reconstruction does not work. However, as a first approximation we have analyzed the
   dominant terms of the  transcendental equation  in
  order to give an approach to the reconstruction of the effective
  potential;  see Fig.\ref{fig2}. Also, we have considered  the extreme limits $\alpha_0/N\ll 1$
  and $\alpha_0/N\gg 1$, in order to find analytical expressions for the
  potential $V(\phi)$. In this approach, we have obtained that in the limit in
  which  $\alpha_0/N\gg 1$, the effective potential coincides with the case in
  which the constant of integration $\beta=0$, where the effective potential
  changes  linearly with the scalar field.

On the other hand, we have explored the possibility of the
reconstruction  in the framework of braneworld inflation,
considering as an  attractor the tensor to scalar ratio in terms of
the number of $e$-foldings i.e., $r=r(N)$. Here we have found
general relation in order to build the effective potential. As a
specific example, we have considered the attractor $r(N)\propto
N^{-2}$. Here, we have obtained that the reconstruction of the
effective potential is given by Eq.(\ref{pote5}). In particular,
considering the limit in which $\alpha_2/\beta_1\gg N$, we have
obtained that the effective potential corresponds to an
exponential potential i.e., $V(\phi)\propto e^{\phi}$;  however
this limit does not work. In the inverse limit, we have found that
the effective
 potential  $V(\phi)=$ constant. Also, utilizing the observables as the scalar
 spectral index and the power spectrum together with the number of $e$-folds, we
 have found a relation between the brane tension and the
 associated parameter $\alpha_1$ to the attractor $r(N)$. Thus, by considering
 that $\alpha_1<252$, in order to obtain $r<0.07$ at $N=60$, we have found an
 upper bound on the brane tension given by $\tau<10^{-12} m_p^4$ and this
 constraint is well corroborated with Planck data, see
 Fig.\ref{fig3}.

We have also found that   in the framework of braneworld inflation,
 the incorporation of the
additional term in  Friedmann's equation affects substantially the
reconstruction of the effective potential $V(\phi)$, considering
the simplest attractors, such as $n_s(N)-1\propto N^{-1}$ or
$r(N)\propto N^{-2}$. In this respect, we have shown that in order
to obtain analytical solutions for the reconstruction of
$V(\phi)$, the attractor $r(N)$, is an adequate methodology to be
considered.

 We conclude with some comments concerning the way to
distinguish the reconstruction in the braneworld and  GR
inflationary models from the methodology used.
 For the famous attractor $n_s-1=-2/N$, we have found that
the reconstruction from $n_s(N)$ in braneworld inflation does not
work unlike  in  GR. Here, we have shown that for a specific case
in which  the integration constants is zero, the reconstruction
from $n_s(N)$ works. On the other hand, by assuming the reconstruction of
braneworld inflation from the attractor $r(N)$, we have been able
to rebuild our model as it occurs in the framework of GR. This suggests that
the version of  reconstruction from $r(N)$ is a suitable ansatz to
be used for the reconstruction of  braneworld inflation.

Finally, in this paper we have not addressed the reconstruction of the braneworld 
model as a  fluid,  considering  an ansatz   on the  effective EoS as a function of the 
number of $e-$folds. We hope to return to this methodology in the near future.

\begin{acknowledgments}
The author thanks to  Manuel Gonzalez-Espinoza and Nelson Videla for useful discussions.
This work was supported
by Proyecto VRIEA-PUCV N$_{0}$. 039.309/2018.
\end{acknowledgments}


\end{document}